\documentclass{ws-ijmpd}

\begin{document}

\markboth{Z. H. Fan \& S. M. Liu}
{Stochastic Acceleration in SNRs}

%
\catchline{}{}{}{}{}
%

\title{Stochastic Electron Acceleration in SNR RX J1713.7-3946}

\author{Zhonghui Fan}

\address{Department of Physics, Yunnan University\\
Kunming 650091, Yunnan, China\\
fanzh@ynu.edu.cn}

\author{Siming Liu}

\address{Key Laboratory of Dark Matter and Space Astronomy, Purple Mountain Observatory, Chinese Academy
of Sciences\\
Nanjing, 210008, P. R. China\\
liusm@pmo.ac.cn}

\maketitle

\begin{history}
\end{history}

\begin{abstract}
Stochastic acceleration of charged particles due to their interactions with
plasma waves may be responsible for producing superthermal
particles in a variety of astrophysical systems. This process can be described as a
diffusion process in the energy space with the Fokker-Planck equation. In this paper, a
time-dependent numerical code is used to solve the reduced Fokker-Planck equation involving only time and energy variables with general forms of the
diffusion coefficients. We also propose a self-similar model for particle acceleration in Sedov explosions and use the TeV SNR RX J1713.7-3946 as an example to demonstrate the model characteristics. Markov Chain Monte Carlo method is utilized to constrain model parameters with observations.
\end{abstract}


\section{Introduction}
\label{intro}

Observations of high-energy emission from many astrophysical sources imply the presence of
significant populations of relativistic particles. Relativistic cosmic rays (CRs) with energies lower than $\sim10^{15}$eV are commonly believed to be accelerated at the shock of supernova remnants (SNRs) in our Galaxy \cite{h05}. Evidence of particle acceleration by shocks of SNRs first comes from radio observations of the remnants, where the radio emission is produced by relativistic electrons via the synchrotron process. The discovery of synchrotron X-ray emission from SNR 1006 reveals acceleration of TeV electrons by these shocks \cite{k95}. However, direct evidence of proton and ion acceleration up to the spectral ``knee'' by these shocks remains elusive.

 Nonthermal distributions are naturally
produced via the Fermi process. The first-order Fermi mechanism corresponds to a convection term in the energy space.
The second-order process, as it was originally conceived by Fermi, involved the
stochastic acceleration (SA) of particles scattering with randomly moving magnetized clouds
\cite{f49}. Later refinements of this idea replaced the magnetized clouds with
magnetohydrodynamic (MHD) waves \cite{m80}.
The second-order stochastic Fermi
process recently finds applications in a wide range of astrophysical environments
(See Refs.~\refcite{pb08}--\refcite{b06} for more details).

One commonly assumed agent of SA is plasma wave turbulence, which is expected to be present in
nonequilibrium conditions of highly magnetized plasmas. The wave-particle interactions in
magnetized plasmas have influences on the dynamics of energetic charged particles and can
play a significant role in the propagation and acceleration of these particles.
Charged particles which spiral along magnetic field lines are accelerated through
resonant interactions with plasma waves. This process can be described as a diffusion process in the energy space through
the Fokker-Planck (F-P) equation with a diffusion coefficient whose magnitude and form
depend on the power spectrum and other characteristics of the plasma turbulence.

The resultant equation in general is complicated and one has to resort to some simplifying approximations. If we are mainly interested in the spatially integrated results, one may obtain a reduced F-P equation with time and energy as the variables. There have been a few studies for analytic solutions of
this reduced F-P equation in simple cases \cite{b06}\cdash\cite{m11}.
However, numerical methods need to be developed to solve more general forms of the reduced equation. Park \& Petrosion (1996) \cite{pp96} studied in detail the numerical methods for the reduced F-P equation.
In this paper, we use their fully implicit  Chang-Cooper method, which is proposed by Chang \& Cooper (1970) \cite{cc70}. Our primary goal in this paper is to present a detailed derivation of the exact time-dependent numerical solution for the reduced F-P diffusion equation. We utilize the Markov Chain Monte Carlo (MCMC) \cite{l11} method to constrain parameters of a model for multi-wavelength observations of TeV SNR RX J1713.7-3946.

\section{Stochastic Electron Acceleration in SNRs}

We consider the generic process of particle acceleration by compressible motions in SNRs
as proposed by  Bykov \& Toptygin (1983) \cite{bt83}. Schlickeiser (1984)
\cite{s84} has shown that the evolution of the overall particle distribution function $N$ can be studied by
solving the diffusion-convection equation in the energy space:
\begin{equation}
\frac{\partial N}{\partial t} = \frac{\partial}{\partial \gamma}
\left[D_{\gamma\gamma}\frac{\partial N}{\partial \gamma}-\left(\frac{2D_{\gamma\gamma}(1+\eta)}{\gamma}
-\frac{\gamma}{3\rm T_{life}}-\dot{\gamma}\right)N\right] -
N{\left(\frac{1}{\rm T_{life}}+\frac{1}{T_{\rm esc}}\right)} + Q(\gamma)\,.
\label{fk04}
\end{equation}
where $\gamma$, $D_{\gamma\gamma}$,  $\rm T_{esc}$ and $\rm T_{life}$ are the Lorentz factor of the particle, the diffusion coefficient in the energy space, the escape time from the acceleration region and the age of the remnant, respectively. $Q$ is a source term and $\dot{\gamma}$ corresponds to the radiative energy loss. Here 
 $\eta$ is related to the first-order Fermi acceleration and the terms related to $\rm T_{life}$ account for the effects of the remnant expansion with an expansion rate of $1/\rm T_{life}$.
The evolution of $N$ is then determined by $D_{\gamma\gamma}$, $\eta$, $T_{\rm esc}$, $\dot{\gamma}$, and $Q$.

For the sake of simplicity, we will assume that $Q$ and $\eta$ are independent of time $t$, and for the study of electron acceleration, we consider the energy loss due to synchrotron and inverse Comptonization of the background radiation \cite{p06}\cdash\cite{m05}.
For stochastic interactions of electrons with the background turbulence, both $D_{\gamma\gamma}$ and $T_{\rm esc}$ are determined by the turbulence properties. We consider a homogeneous and isotropic turbulence with a Kolmogorov spectrum, then the characteristic length of the magnetic field in a high $\beta$ plasma is given by $L = C_1(v_{\rm A}/U)^3 R$, where $R$, $U$, and $v_{\rm A}$ are the radius of the remnant, the shock speed, and the Alfv\'{e}n speed, respectively \cite{f10b}. $C_1$ is a dimensionless constant.
The spatial diffusion coefficient of relativistic particles is then given by
\begin{equation}
{\cal X} = cL/3 + \alpha {\gamma m_ec^3/(3 eB)}
\label{diff1}
\end{equation}
where we have assumed super-Bohm diffusion (with the dimensionless constant $\alpha>1$) for electrons with a gyro-radius $r_g=\gamma m_ec^2/eB$ greater than $L$, and $c$, $m_e$, $e$, and $B$ are the speed of light, electron mass and charge, and the magnetic field, respectively. Then we have
\begin{eqnarray}
D_{\gamma\gamma}&=&\gamma^2U^2/9{\cal X}\,,\\
T_{\rm esc}&=&R^2/{\cal X}\,.
\end{eqnarray}
To have a self-similar solution, $T_{\rm esc}$ needs to scale with the age of the remnant $T_{\rm life}$. For the Sedov solution of SNR evolution with $R\propto T_{\rm life}^{2/5}$ and $U\propto T_{\rm life}^{-3/5}$, this implies that ${\cal X}\propto RU$, which, in combination with equation (\ref{diff1}), gives $v_{\rm A}\propto U^{4/3}\propto T_{\rm life}^{-4/5}$ and $B\propto T_{\rm life}^{-4/5}$.

\begin{figure}[h]
\includegraphics[height=.25\textheight,width=0.49\textwidth]{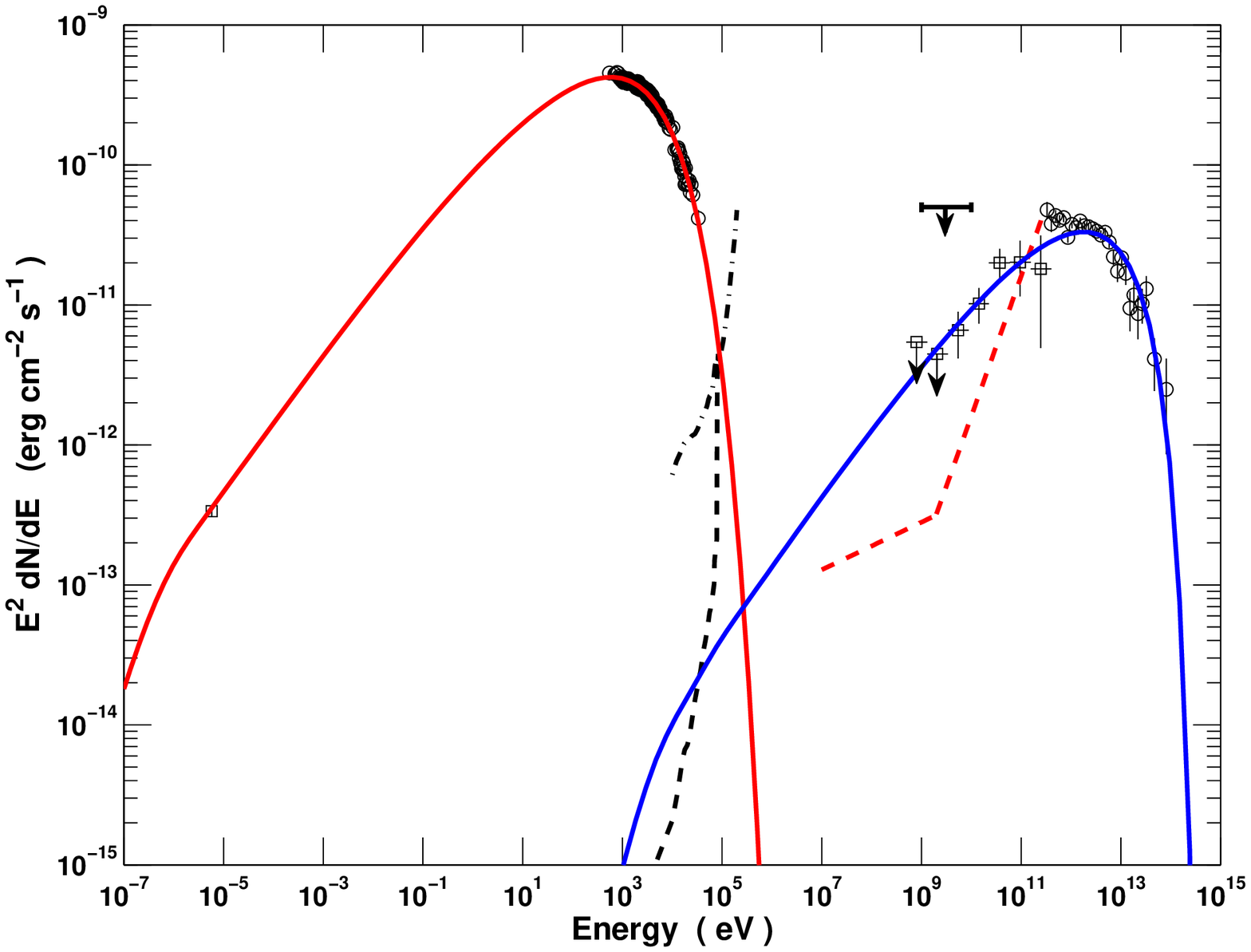}
\includegraphics[height=.25\textheight,width=0.49\textwidth]{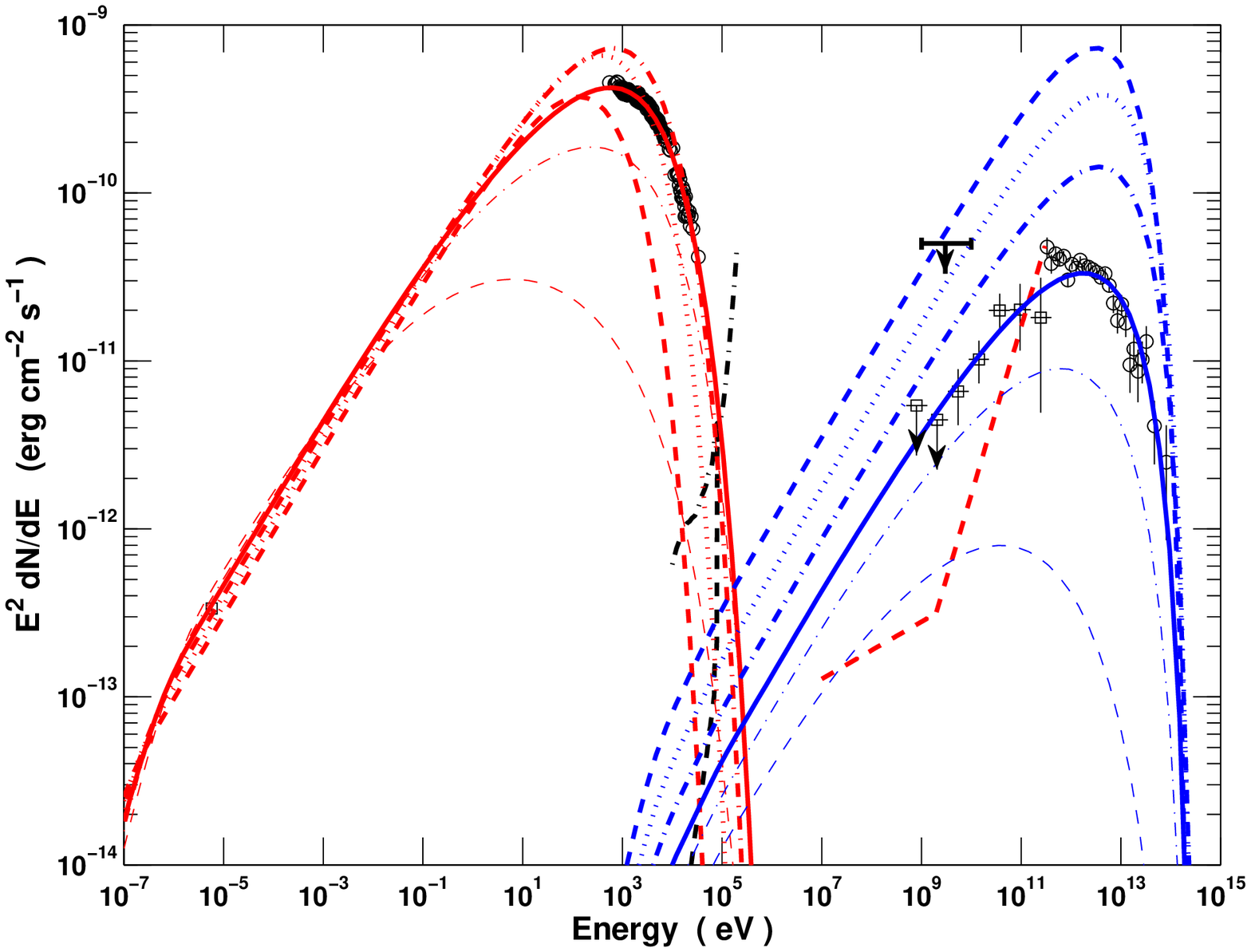}
\caption{The left panel shows the SED of the self-similar solution proposed in this paper at 1600 years;
The right panel shows the evolution of SED: 500 yr (dashed line), 1000 yr
(dotted-dashed line), 1600 yr (thick line), 3200 yr (thick dotted-dashed
line), 6400 yr (thick dotted line) and 12800 yr (thick dashed line).
\label{fig}}
\end{figure}

\begin{figure}[ht]
\centering
\includegraphics[height=.35\textheight,width=0.9\textwidth]{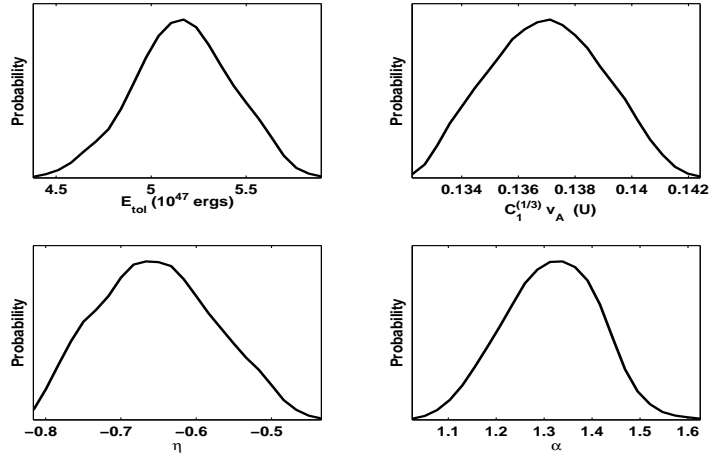}
\caption{Probability distribution function of model parameters normalized at
the peak value.
\label{fig3}}
\end{figure}

For the shell Type TeV SNR RX J1713.7-3946, we adopt $T_{\rm life}=1600$ yr, $R = 10(T_{\rm life}/1600{\rm yr})^{2/5}$ pc,  $U = 3500(T_{\rm life}/1600{\rm yr})^{-3/5}$ km/s, $B = 12.0(T_{\rm life}/1600{\rm yr})^{-4/5}\,\mu$G, and assume that the electron acceleration starts at $70$ years after the supernova explosion \cite{c84} and the electrons are injected at the proton rest mass energy with $\gamma = 1836$. Kinetic plasma waves can accelerate electrons to the proton rest mass energy efficiently at even lower energies \cite{pl04}\cdash\cite{l06}. Then by fitting the emission spectrum (Fig. \ref{fig}), we have $\eta=-0.67$, $\alpha= 1.3$,  $C_1^{1/3}v_{\rm A}=0.137 U$, and the total energy of electrons with $E_{\rm tol} =5.2\times 10^{47}$ ergs. The corresponding reduced $\chi^2$ of the best fit is $1.83$.
In Figure \ref{fig}, the radio datum is from Acero et al. (2009) \cite{a2009}, other data are from Fan et al. (2010) \cite{f10a}. The right panel of Figure \ref{fig} shows the model predicted spectral evolution. Since the injection rate is constant and the magnetic field is proportional to $T_{\rm life}^{-4/5}$, the radio luminosity has a very weak dependence on time while the GeV luminosity increases roughly linearly with time. The spectral hardening in the TeV band in the early phase of the evolution is closely related to the electron acceleration process.
We notice that the self-similar solution may not be valid after 10000 years since the magnetic field falls below $3 \mu$G, which is considered a typical value of the interstellar medium.
Figure \ref{fig3} shows the probability distribution function of model parameters given by the MCMC method. It is interesting to note that the value of $\alpha$ of the best fit is very close to $1$ and $\eta$ is negative, implying a first-order energy loss process.

\section{Summary and Conclusions}
\label{dis}

The F-P equation in general does not have analytical solutions for the diffusion
coefficient in the SA theory. To obtain the electron distribution from
the F-P equation, one may resort to some simplifying approximations
involving only the time and energy variables.
In the paper, we develop a numerical code to solve this reduced F-P equation,
which is applicable for arbitrary forms of the diffusion coefficient.

Recent observations of SNR RX J1713.7-3946 made with the {\it Fermi} space telescope show that the spectrum of this source in the GeV band is very hard with a power-law photon index of $\Gamma=1.5\pm0.1$, which is difficult to accommodate in a hadronic scenario, but agrees well with the IC origin of $\gamma$-rays in the leptonic scenario \cite{a11}. On the other hand, by considering the potentially high inhomogeneity of the shocked interstellar medium (ISM), \cite{i11} argue that a hadronic model may still explain the Fermi observation.
The nature of the TeV emission from SNR is an issue still open to investigation.
In the work, we show that the SA of electrons by turbulent plasma waves in the
shock downstream might naturally explain these observations \cite{li11}.
Future observations with the {{\it HXMT}} and {{\it NuSTAR}} can test the model.

\section*{Acknowledgments}

This work is supported by the National Science Foundation of China
(grants 10963004, 11143007, 11173064), Yunnan Provincial Science Foundation of China (grant 2008CD061) and SRFDP (grant 20095301120006). We acknowledge the use of CosRayMC \cite{l11} adapted from COSMOMC \cite{l02}.


\end{document}